\documentclass[pra,aps,twocolumn]{revtex4}
\usepackage{graphicx}

\begin{document}

\title{Kapitza homogenization of deep gratings for designing dielectric metamaterials}

\author{Carlo Rizza$^{1,3}$, Alessandro Ciattoni$^1$, }
\affiliation{$^1$Dipartimento di Fisica e Matematica, Universit\`a dell'Insubria, Via Valleggio 11, 22100 Como, Italy} \affiliation{$^2$Consiglio
Nazionale delle Ricerche, CNR-SPIN 67100 L'Aquila, Italy and Dipartimento di Fisica,
             Universit\`a dell'Aquila, 67100 L'Aquila, Italy}

\begin{abstract}
We theoretically investigate the homogenization of the dielectric response to transverse electric waves of a transverse grating characterized by the
Kapitza condition, i.e. the permittivity is rapidly modulated with a modulation depth scaling as the large wavelength to modulation period ratio. We show
that the resulting effective dielectric permittivity, in addition to the standard average of the underlying dielectric profile, has a further contribution
arising from the fast and deep dielectric modulation. Such contribution turns out to be comparable with the other one and hence can provide an additional
way for designing dielectric metamaterials. As an example, we discuss an effective metal-to-dielectric transition produced by the Kapitza contribution
obtained by changing the grating depth.
\end{abstract}

\maketitle

Materials patterned with a periodic dielectric permittivity are a major platform for achieving electromagnetic radiation steering and manipulation
\cite{Joannopoulos_1}. If the modulation period is much smaller than the wavelength, the periodic medium displays an homogeneous dielectric response in
turn described by effective material parameters \cite{Cai_1}. In the metamaterial context, subwavelength layered metal-dielectric structures have
attracted a great attention due to the number of exotic electromagnetic phenomena they host resulting from the possibility of tailoring both the magnitude
and the sign of the effective permittivity. Leading examples of such unusual phenomena are subwavelength imaging \cite{Pendry_1,Belov_1}, negative
refraction \cite{Scalora_1}, hyperbolic response \cite{Jacob_1}, optical non-locality \cite{Elser_1,Orlov_1}, transverse power flow reversing
\cite{Ciattoni_1} and transmissivity directional hysteresis \cite{Ciattoni_2} in extreme nonlinear metamaterials. Generally, the response of a dielectric
stratified composite is theoretically described by the standard effective medium theory \cite{Cai_1} according to which the ordinary and extraordinary
effective permittivities are the weighted arithmetic and the harmonic means of the underlying layer permittivities, respectively. On the other hand, in
Ref.\cite{Elser_1}, Elser {\it et al.} proposed an improvement of such standard effective medium approach by showing that the optical response of a
realistic multi-layered nano-composite can be strongly affected by non-local effects. Recently, Rizza {\it et al.} \cite{Rizza_1} have suggested that a
novel propagation regime, not encompassed by the standard effective medium predictions, occurs if a Transverse Magnetic (TM) wave propagates along the
modulation direction of a very deep and subwavelength grating. Here the large grating depth plays a fundamental role and, more precisely, if such depth
scales as the large ratio between wavelength and the grating period (\emph{Kapitza condition}), the authors show that a different mechanism supporting
diffractionless propagation occurs which is not significantly hampered by medium losses.

In this Letter, we theoretically investigate the propagation of Transverse Electric (TE) waves in a realistic transverse dielectric grating fulfilling the
Kapitza condition. We show that the rapid grating with a large depth entails a medium homogenization which can not be described by the standard effective
medium theory. Accordingly, in addition to the average of the underlying permittivity profile, the ensuing dielectric effective permittivity has a further
contribution which is generally not a small perturbation. In addition, we study the transition from the standard effective medium regime to the Kapitza
one by keeping fixed the grating period and gradually increasing the modulation depth and we show that, in the considered example, the dielectric Kapitza
correction turns the metallic medium behavior into a dielectric one. The considered electromagnetic situation has a stringent analogy with the classical
mechanical motion of a particle subjected both to a conservative field (described by a potential energy $U$ with a characteristic variation time $T$) and
to a rapidly varying force (oscillating with a frequency $\Omega$) \cite{Landau_1,Kapitza_1}. In this case, the condition $\Omega T>>1$ entails a
separation between the slow and fast dynamical scales and, in complete analogy with the dielectric effective permittivity of the Kapitza medium, the
effective potential energy driving the slow motion is the superposition of the average of $U$ and a contribution resulting from the underlying fast
dynamics. Such approach has been extended in several physical context and in particular in optics for achieving nonlinearity management
\cite{Hasegawa_1,Ciattoni_4}.

Let us consider a monochromatic TE electromagnetic field ${\bf E}(x,z)=Re\left\{E(x,z) \exp \left[-i t \left(2\pi c \right)/\lambda\right]\right\}
\hat{\bf e}_y$ (of wavelength $\lambda$) propagating in a periodic medium and satisfying the Helmholtz equation
\begin{equation}
\label{Helm} \nabla^2 E+k_0^2 \epsilon(x) E=0,
\end{equation}
where the period of the dielectric permittivity $\epsilon(x)$ is $\Lambda$ and $k_0=2 \pi/\lambda$ ($\nabla=\hat{\bf e}_x\partial_x+\hat{\bf
e}_z\partial_z$). Specifically, we consider a medium whose relative dielectric permittivity admits the Fourier series expansion
\begin{equation}
\label{Fourier} \epsilon=\epsilon_m+\sum_{n\neq 0}\left(a_n + \frac{b_n}{\eta} \right) \exp \left(i n \frac{k_0}{\eta} x\right),
\end{equation}
and which satisfies the Kapitza condition, i.e. the spatial grating modulation is very fast ($\eta=\Lambda/\lambda\ll 1$) and the grating amplitude is
very large, it scaling as the inverse of $\eta$. Following the Kapitza approach \cite{Landau_1} and exploiting the multiscale asymptotic expansion, we
look for a solution of Eq.(\ref{Helm}) as
\begin{equation}
\label{sol} E(x,X,z)=\bar{E}(x,z)+\eta \sum_{n\neq 0} \tilde{E}_n (x,z) \exp \left(i n k_0 X\right),
\end{equation}
where $X=x/\eta$ is a fast coordinate, whereas $\bar{E}$ and the last term in the RHS are the slowly and rapidly varying parts of the electric field,
respectively. Substituting the ansatz of Eq.(\ref{sol}) into Eq.(\ref{Helm}) and selecting the terms with the same spatial frequencies (hence dropping the
fast scale), we obtain the coupled equations
\begin{eqnarray}
\label{Helm-scale}
&&\nabla^2 \bar{E}+k_0^2 \epsilon_m \bar{E}+k_0^2 \sum_{n\neq 0} \left(b_{-n}+\eta a_{-n}\right)\tilde{E}_n =0, \nonumber \\
&&\left(b_n \bar{E}-n^2 \tilde{E}_n \right) + O(\eta) = 0
\end{eqnarray}
where $n\neq0$ and $O(\eta)/\eta$ is a finite complex number in the limit $\eta \rightarrow 0$. In the asymptotical limit $\eta \rightarrow 0$, the second
of Eqs.(\ref{Helm-scale}) turns into a simple algebraic equation, i.e. $\tilde{E}_n=b_n/n^2 \bar{E}$ so that the first of Eqs.(\ref{Helm-scale}) yields,
for $\eta \rightarrow 0$, $\nabla^2 \bar{E}+k_0^2 \epsilon_{\mathrm{eff}} \bar{E}=0$, where
\begin{equation}
\label{eff} \epsilon_{\mathrm{eff}}=\epsilon_m+\sum_{n\neq 0} \frac{b_{-n} b_n}{n^2}.
\end{equation}
Note that, in analogy with the mechanical situation, the obtained effective dielectric permittivity of Eq.(\ref{eff}), in addition to the weighted
arithmetic mean of the permittivity profile $\epsilon_m$ (as predicted by the standard effective medium theory) has a further contribution produced by the
small (but fast) component of the electric field $\eta \tilde{E}_n$, and this term can significantly affect the dynamics of the slowly average and leading
order field $\bar{E}$.

In order to check the goodness of the multiscale asymptotic expansion and to discuss the impact of the actual small value of $\eta$, we consider TE waves
impinging onto a Kapitza slab of thickness $L$ as reported in Fig.1.
\begin{figure}
\center
\includegraphics*[width=0.48\textwidth]{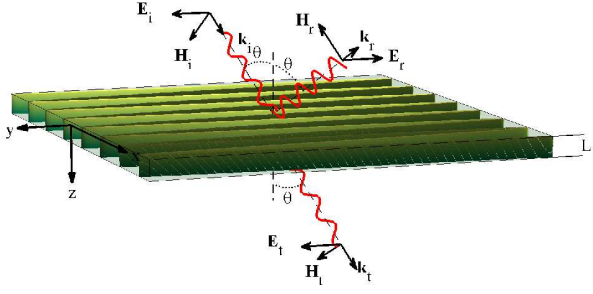}
\caption{(Color on-line) Sketch of a TE plane wave impinging onto a Kapitza slab of thickness $L$ with an incidence angle $\theta$.}
\end{figure}
Specifically, as a theoretical benchmark, we assume the slab grating to be described by the permittivity
\begin{equation}
\label{cos} \epsilon=\epsilon_m+\delta \epsilon \cos\left[(2 \pi/\Lambda) x\right].
\end{equation}
We have numerically evaluated the transmissivity and the reflectivity of the slab by resorting to $2$D full-wave simulations performed with the comsol RF
module \cite{comsol}. In our simulations, we have considered two vacuum layers placed at the facets of the slab for providing external excitation and
hence for evaluating the scattering parameters (transmissivity and reflectivity). We analyzed only one period along the $x$-axis imposing Floquet phase
shift boundary condition on the sides of the unit cell. Furthermore, we have used matched boundary conditions at the entrance and exit facets (orthogonal
to the $z$-axis) of the integration domain \cite{comsol}. In Fig.2, we report the numerical transmissivity $T=|{\bf E}_\mathrm{t}|^2/|{\bf
E}_\mathrm{i}|^2$ and reflectivity $R=|{\bf E}_\mathrm{r}|^2/|{\bf E}_\mathrm{i}|^2$ labelled with square and circle solid lines, respectively (see Fig.1
for the definitions of the incident ${\bf E}_\mathrm{i}$, reflected ${\bf E}_\mathrm{r}$, and transmitted ${\bf E}_\mathrm{t}$ field amplitudes,
respectively) with $\lambda=100$ $\mu$m, $\Lambda=\lambda/10$, $\epsilon_m=-0.25+i 0.0125$, $Im(\delta \epsilon)=0.01$ and $L=300$ $\mu$m. We plot $T$ and
$R$ as functions of the grating amplitude $Re(\delta \epsilon)$ for normally impinging waves (panel (a)) and as functions of the incidence angle for
$Re(\delta \epsilon)=10$ (panel (b)). On the other hand, using the results of the multiscale approach, we have analytically evaluated the transmissivity
$T_{\mathrm{eff}}$ and the reflectivity $R_{\mathrm{eff}}$ by regarding the slab as homogeneous with the effective permittivity
$\epsilon_{\mathrm{eff}}=\epsilon_m + \left[\eta Re\left(\delta \epsilon\right) \right]^2/2$ evaluated by using Eq.(\ref{eff}) for the profile of the
dielectric constant of Eq.(\ref{cos}). For comparison purposes, in Fig.2(a) and Fig.2(b), we also plot $T_{\mathrm{eff}}$ (dashed lines) and
$R_{\mathrm{eff}}$ (dot-dashed lines).
\begin{figure}
\center
\includegraphics*[width=0.48\textwidth]{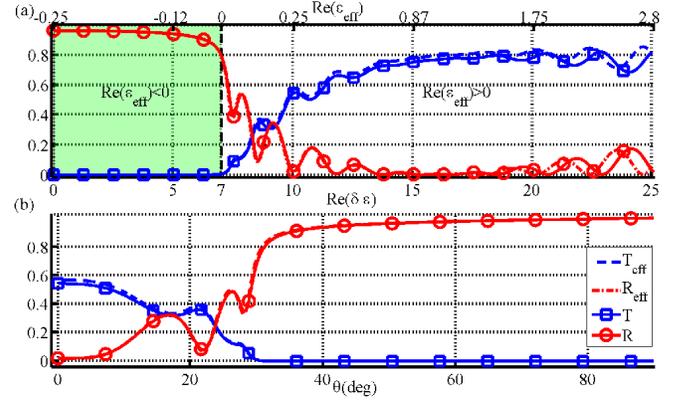}
\caption{Comparison between numerical slab transmissivity $T$ and reflectivity $R$ and corresponding Kapitza quantities $T_{\mathrm{eff}}$ and
$R_{\mathrm{eff}}$. (a) normally imping waves for different grating depths; (b) inclined waves for $Re(\delta \epsilon)=10$.}
\end{figure}
It is evident that the agreement between the numerical simulations and the predictions of the Kapitza approach is quite remarkable even for the chosen not
so small ratio $\Lambda/\lambda = 1/10$. From Fig. 2(a) we note that, for small values of $Re\left(\delta \epsilon\right)$, both $T$ and $R$ slightly
depend on the grating depth as predicted by the standard effective medium theory. On the other hand, a threshold exists ($Re\left(\delta \epsilon\right)
\simeq 7$) after which $T$ and $R$ show a marked dependence on $Re\left(\delta \epsilon\right)$ proving that the Kapitza regime has been reached. In
addition, Fig.2(a) also shows that the Kapitza contribution to $\epsilon_{\mathrm{eff}}$ can generally play an important role since, in the considered
situation, by varying the grating amplitude $Re\left(\delta \epsilon\right)$, we observe a metal-to-dielectric transition: the real part of effective
dielectric permittivity is negative and positive in the region $0<Re\left(\delta \epsilon\right)<7$ (shadow area) and in the region where $Re\left(\delta
\epsilon\right)>7$, respectively. Figure 2(b) shows that the Kapitza approach holds for all the incident angles. Note that, by comparing Eqs.(\ref{cos})
and (\ref{Fourier}) for the situation reported in Fig.2(b), we have $a_{\pm 1}=i 0.005$, $b_{\pm 1}=0.5$ ($a_n=b_n=0$ for $|n| \neq 1$) so that the
Kapitza requirement of the grating depth scaling as the grating period to wavelength ratio is numerically fulfilled ($b_{\pm 1}$ being close to one).
\begin{figure}
\center
\includegraphics*[width=0.48\textwidth]{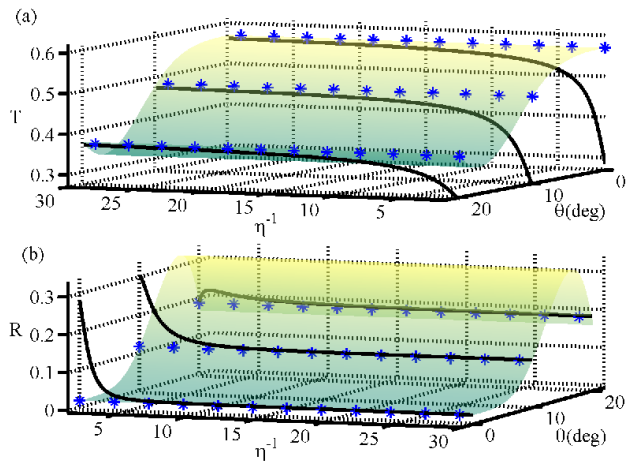}
\caption{Comparison between numerical transmissivity (a) and reflectivity (b) (solid lines) and those predicted by Kapitza approach (stars) as functions
of $\theta$ and $1/\eta$. The semitransparent surface interpolating the results of the Kapitza approach is reported for clarity purposes.}
\end{figure}
In Fig.3, we plot the numerical transmissivity (panel (a)) and reflectivity (panel (b)) along with those predicted by Kapitza approach letting both the
grating depth and period to vary with the constraint $Re(\delta \epsilon)=\eta^{-1}=\lambda/\Lambda$. Here the Kapitza requirement of the grating depth
scaling as the grating period to wavelength ratio is evidently numerically fulfilled. Both $T$ and $R$ have been evaluated for different values of $\eta$
and $\theta$ in the case $\lambda=100$ $\mu$m, $\epsilon_m=-0.25+i 0.0125$, $Im(\delta \epsilon)=0.01$, $L=300$ $\mu$m. From Fig.3, we note that the
discrepancy between the numerical and Kapitza predictions is significant for $\eta^{-1}<5$ and it is negligible for increasing value of $\eta^{-1}$, in
agreement with the above discussed multiscale asymptotic analysis.

In conclusion, we have studied the homogenization of the electromagnetic response of a transverse dielectric grating satisfying the Kapitza condition,
i.e. the ratio between the wavelength and the grating period is much larger than one and the grating depth scales as this ratio. We have showed that the
regime is described by a novel effective medium theory predicting for the effective permittivity a Kapitza contribution to the average permittivity which
is generally not a small perturbation. We expect our approach to offer new avenues for the tailoring of metamaterial properties and specifically for the
design of epsilon-near-zero metamaterials \cite{Engheta}.

This research has been funded by the Italian Ministry of Research (MIUR) through the "Futuro in Ricerca" FIRB-grant PHOCOS - RBFR08E7VA.

\end{document}